\documentclass[11pt]{article}
\usepackage[margin=1in]{geometry}
\usepackage[skip=10pt plus1pt]{parskip}
\usepackage[utf8]{inputenc} 
\usepackage[T1]{fontenc}    
\usepackage{url}            
\usepackage{booktabs}       
\usepackage{nicefrac}       
\usepackage{microtype}      
\usepackage{psfrag}
\usepackage{array}
\usepackage{booktabs}
\usepackage{siunitx}
\usepackage{framed,multirow}

\usepackage[usenames, dvipsnames]{xcolor}
\definecolor{myBlue}{rgb}{0.2, 0.2, 0.6}
\definecolor{myOrange}{rgb}{0.82, 0.41, 0.12}

\usepackage{csquotes}
\usepackage{caption}
\usepackage{subfigure}
\usepackage{wrapfig}
\usepackage{amsmath}
\usepackage{amsbsy}
\usepackage{psfrag}
\usepackage{pstool}
\usepackage{graphicx}
\usepackage{algorithm}
\usepackage{algpseudocode}
\usepackage{lineno}
\usepackage{hyperref}
\usepackage{utfsym}
\usepackage{wrapfig}

\hypersetup{breaklinks = true,
	      colorlinks = true,
	      linkcolor  = blue,
            urlcolor   = blue,
	      citecolor  = blue}

\modulolinenumbers[5]
\usepackage{tikz}
\usetikzlibrary{plotmarks}

\usepackage{multicol}
\usepackage{authblk}

\author[1]{Simone~Silvestri}
\author[1]{Gregory L.~Wagner}
\author[1]{Christopher~Hill}
\author[2]{Matin~Raayai~Ardakani}
\author[3]{Johannes~Blaschke}
\author[1]{Jean-MichelvCampin}
\author[1]{Valentin~Churavy}
\author[4]{Navid C.~Constantinou}
\author[1]{Alan~Edelman}
\author[1]{John~Marshall}
\author[1]{Ali~Ramadhan}
\author[1]{Andre~Souza}
\author[1]{Raffaele~Ferrari}
\affil[1]{Massachusetts Institute of Technology, Cambridge, MA, USA}
\affil[2]{Northeastern University, Boston, MA, USA}
\affil[3]{Lawrence Berkeley National Laboratory, Berkeley, CA, USA}
\affil[4]{Australian National University, Canberra, ACT, Australia}
\title{Oceananigans.jl: A Julia library that achieves breakthrough resolution, memory and energy efficiency in global ocean simulations}
\date{March 2023}

\begin{document}

\maketitle

\begin{abstract}

\noindent Climate models must simulate hundreds of future scenarios for hundreds of years at coarse resolutions, and a handful of high resolution decadal simulations to resolve localized extreme events.
Using Oceananigans.jl, written from scratch in Julia, we report several achievements: First, a global ocean simulation with breakthrough \textit{horizontal resolution} --- 488m --- reaching 15 simulated days per day (0.04 simulated years per day; SYPD).
Second, Oceananigans simulates the global ocean at 488m with breakthrough \textit{memory efficiency} on just 768 Nvidia A100 GPUs, a fraction of the resources available on current and upcoming exascale supercomputers.
Third, and arguably most significant for climate modeling, Oceananigans achieves breakthrough \textit{energy efficiency} reaching 0.95~SYPD at 1.7~km on 576~A100s and 9.9~SYPD at 10~km on 68~A100s --- the latter representing the highest horizontal resolutions employed by current IPCC-class ocean models.
Routine climate simulations with 10 km ocean components are within reach.

\end{abstract}

\section{Justification}

Oceananigans.jl --- a new ocean model written from scratch in Julia --- achieves ocean simulations with breakthrough resolution, memory and energy efficiency, realizing 0.041 simulated years per day (SYPD) at 488 m on 768 Nvidia A100s, 0.95 SYPD at 1 km on 576 A100s, and 9.9 SYPD at 10 km on 68 A100s.

\section{Performance Attributes}

\begin{center}
\begin{tabular}{c|l}
\hline
\hline
Categories & Scalability, time-to-solution, energy-to-solution. \\
Type of method & Fully explicit with sub-cycling. \\
Results basis & Whole application excluding I/O. \\
Numerical precision & Both 64- and 32-bit cases measured. \\
System scale & Results measured on full-scale systems. \\
Measurement mechanism & Timers, memory used and energy used. \\
\hline
\hline
\end{tabular}
\end{center}

\section{Overview of the Problem}
\label{sec:overview}

Climate models are essential for predicting where, when, and how climate change threatens Earth's ecosystems and human civilization.
But current climate models, which capture only the broadest aspects of global warming, fall far short of providing the needed accuracy and granularity required to design and implement costly adaptation and mitigation strategies~\cite{Fiedler21a}.
Significant reduction of the uncertainty of climate predictions is potentially worth trillions of dollars~\cite{hope}.

Climate models simulate the three-dimensional fluid dynamics, thermodynamics, chemistry, and biology of the atmosphere, ocean, and land to predict the hydrological cycle, carbon cycle and the net energy imbalance of the Earth system.
While typical climate models use coarse resolutions of 25-100 km to simulate the numerous climate scenarios required by the Intergovernmental Panel on Climate Change (IPCC)~\cite{ipcc}, a handful of state-of-the-art climate simulations have been performed at higher resolutions of O(10 km) at astronomical expense.
At either resolution there are many processes, such as clouds and ocean turbulence, that cannot be explicitly simulated and are instead approximated by empirical formulae called \textit{parameterizations}.
Biases due to inadequate parameterizations dominate the uncertainty of climate predictions over the next few decades~\cite{palmer, kendon2014heavier}.

The prevailing strategy to reduce climate model uncertainty is to refine model resolution as much as possible \cite{palmer}.
For example, at horizontal resolutions of 1~km a substantial fraction of atmospheric convection and ocean turbulence are explicitly modeled by Newton's laws of motion, greatly reducing the impact of parameterizations~\cite{palmer}.
High-resolution climate modeling is further required to make predictions for specific regions, providing information for local decision makers on adaptation and mitigation~\cite{Fiedler21a}.

Yet the ``resolution strategy'' is fundamentally limited: even at 1~km resolution many climate-relevant physical processes remain unresolved~\cite{wedi2020baseline}.
Worse, processes such as sea ice dynamics, biology, or cloud-aerosol interaction will never be resolved because accurate macroscopic laws do not exist.
Absent theoretical breakthroughs, such ``irreducible'' uncertainties can be addressed only by leveraging Earth system observations through advances in data assimilation and machine learning~\cite{schneider}. 
Data-driven optimization of climate models requires \textit{ensembles} of climate predictions, rather than single predictions at the highest affordable resolution.
Ensembles of simulations are also required to explore emission scenarios and to estimate the impact of initial condition uncertainty and internal variability.

Consequently, reducing the uncertainty of climate predictions demands not \textit{just} higher resolution, but \textit{more efficient resource utilization} to enable hundreds to thousands of relatively high-resolution simulations.
As an example, we consider the computational requirements to enable 100-simulation ensembles using all 37,888 AMD MI250 GPUs of the Frontier exascale supercomputer: completing an ensemble of 300-year simulations (200 years of spin-up + 100 years of prediction) within one month of wall clock time requires a climate model that can achieve 10 simulated years per day (SYPD) using 378 GPUs, or 1/100th of Frontier's resources.
Disruptive progress on climate modeling requires not just scalable performance for a single, high-resolution simulation, but advances in \textit{efficiency} to meet this ensemble-based ``10 per 100th'' benchmark \cite{harnessingai}.

Our submission uses the ocean component of a new climate model being developed by the Climate Modeling Alliance~\cite{clima}. 
The ocean contributes key uncertainty to climate predictions due to its prominent role in the Earth system's heat and carbon cycles.
At 10~km resolutions, where ocean model uncertainties are significantly reduced, the ocean often is the most expensive climate model component~\cite{nemo-highres}.
This calls for a step change in ocean model performance.

\section{Current State of the Art}

We are aware of only three global ocean simulations that have achieved resolutions finer than 5 kilometers --- all at tremendous computational expense.
In 2014, MITgcm \cite{marshall1997finite} was used to perform the one year, tidal-forced ice-ocean simulation ``LLC4320'' \cite{llc4320}, which exhibits 2.2 km horizontal resolution with 90 vertical levels.
LLC4320 achieved 0.047 simulated years per day (SYPD) using 70,000 cores of the NASA Pleiades system.

\begin{wrapfigure}{r}{0.6\textwidth}
  \begin{center}
    \includegraphics[width=0.6\textwidth]{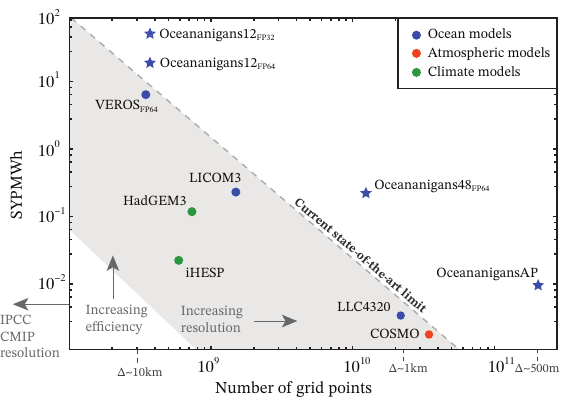}
  \end{center}
  \small
  \caption{\footnotesize Simulated years computed by a megawatt-hour of energy (SWPMWh) versus number of grid points for state-of-the-art atmosphere and ocean models.
  Stars show the performance of our ocean model in a realistic and ``aqua planet'' (AP) setup.}
  \label{fig:sypmwh}
\end{wrapfigure}
\normalsize

FIO-COM32 \cite{fiocom32} ran at $\sim$2.5 km (1/32${}^{\rm{nd}}$ degree) horizontal resolution with 90 vertical levels for 3.5 years.
\cite{licom3} ported LICOM3 to GPUs to realize 0.51 SYPD at 1/20${}^{\rm{th}}$ horizontal resolution with 60 vertical levels using 384 MI50 AMD GPUs, and further managed to scale to 26200 MI50s with strong scaling efficiency of 8\%.

The largest ocean simulations used in current IPCC-class climate models, which typically require faster-time-to-solution to support longer simulations, have horizontal resolutions of roughly 10~km.
\cite{chassignet2020impact} describes output from four 60-year ocean simulations following the OMIP-2 protocol with 8~km (1/12${}^{\rm{th}}$ degree), 10~km, and two with 11~km (1/10${}^{\rm{th}}$ degree).
\cite{century} report a 110-year simulation at 10 km (1/10${}^{\rm{th}}$ degree) horizontal resolution, the longest high resolution OMIP-2-style run. 
Some of the highest resolution climate models are the iHESP CESM-based model with 25km-10km atmosphere-ocean resolution \cite{zhang2020optimizing}, achieving 3.4 SYPD, and the 50km-10km HadGEM3-GC3.1 submission to HighResMIP \cite{annexipcc, highresmip}, achieving 0.4 SYPD.

At 3.4 SYPD, the iHESP CESM achieves sufficient time-to-solution for hundreds to thousands of years of simulated climate.
But such a simulation is purchased for a high price, requiring the 40\% of the Sunway TaihuLight supercomputer \cite{zhang2020optimizing} and 4 million cores consuming 6 MW for hundreds of days of wall clock time.
Enabling the large ensembles of high-resolution simulations needed to improve climate prediction requires both performance at scale as well as efficient \textit{resource utilization}.

Figure~\ref{fig:sypmwh} plots simulated years per \textit{mega-watt-hour} (SYPMWh) against resolution for state-of-the-art ocean models.
The SYPMWh metric encodes the efficiency requirement needed to make progress on climate uncertainty with next-generation climate models: in particular, we require both higher-resolution models (moving rightwards in figure~\ref{fig:sypmwh}) as well as more efficient models (moving upwards in figure~\ref{fig:sypmwh}).
For completeness, we report SYPMWh also for two GPU-based models: Veros~\cite{veros}, an ocean model, and COSMO \cite{cosmo}, an atmospheric model.
The present nomination is shown with stars from whence we see significant performance gains compared to the existing state-of-the art.

\section{Innovations}

Our achievement is three-fold: first, using new software written in the Julia programming language called Oceananigans.jl~\cite{ramadhan2020oceananigans}, we report a near-global ocean simulation with highest-ever horizontal resolution (488 m) reaching 15 simulated days per day (0.04~SYPD).
Second, Oceananigans performs this simulation with breakthrough \textit{memory efficiency} on just 768 NVidia A100 GPUs, and thus a fraction of the available resources on current and upcoming exascale supercomputers.
Third, and arguably most important, Oceananigans achieves breakthrough \textit{energy efficiency}, simulating the global ocean at 0.95~SYPD with 1.7~km resolution on 576~A100s, and at 10 km --- the highest horizontal resolution employed by an IPCC-class ocean model --- achieving 9.9~SYPD on 68 Nvidia A100s.
This final milestone proves the feasibility of \textit{routine} climate simulations with 10~km ocean components, a crucial resolution threshold at which ocean macroturbulence (the most energetic ocean motions with scales between 10--100 km) is fully resolved.

We attribute these achievements first and foremost to a high-risk, high-reward strategy to develop a new ocean model from scratch in Julia with a specific focus on GPU performance and memory efficiency.
Additional crucial ingredients include advances in numerical methods for finite volume fluid dynamics on the sphere and a novel optimization for simulating ocean free surface dynamics that achieves unprecedented GPU scalability.

\subsection{Starting from scratch with Julia}

Oceananigans.jl is an open-source library for ocean-flavored fluid dynamics written from scratch in Julia~\cite{Bezanson2017-ca}.
Julia is a dynamic high-level programming language that leverages Just-In-Time (JIT) compilation and LLVM~\cite{Lattner2004-hw} to achieve performance competitive with traditional HPC languages like C or Fortran. 
Julia has gathered interest as potential language for HPC~\cite{Godoy2023-fx,Churavy2023-ja,Giordano2022-hi,Hunold2020-wm,Lin2021-rk} and provides easy integration with MPI~\cite{Forum1994-nq,Byrne2021-hl}.
Most of Oceananigans.jl software is hardware-agnostic through the Julia package KernelAbstractions.jl~\cite{Churavy2023-ja}, which enables performance portability targeting CPUs and different GPU vendors using the JuliaGPU~\cite{Besard2017-it,Besard2019-jz} software stack, similar to the capabilities provided by Kokkos~\cite{trott2021kokkos}, OCCA~\cite{medina2014occa}, and HIP~\cite{hip}.

To our knowledge, Oceananigans is the first ocean model written from scratch for GPUs, rather than ported from existing CPU code.
Starting from scratch and using the Julia programming language allowed us to rethink the typical patterns used in ocean and atmosphere dynamical cores.
In particular, we developed a system of composable atomic operators that leverages Julia's functional programming paradigm and effective inlining capabilities to recursively construct large expression trees for calculus on staggered finite volume grids.
Using this composable operator system, we fuse the entire tendency computation for each prognostic variable into a single compute-heavy kernel, each of which depends on only two intermediate diagnostic variables representing hydrostatic pressure and vertical diffusivity (which is treated implicitly using a predictor-corrector method).

Such a high degree of abstraction yields a number of innovations: first, kernel fusion maximizes efficiency on GPUs.
Second, almost all intermediate quantities are computed on-the-fly, so that Oceananigans is extremely memory efficient and can perform global ocean simulations at resolutions up to 1/4${}^{\rm{th}}$ degree on a single Nvidia V100.
Finally, because all compute-heavy kernels rely on a single ``tendency kernel function'' applied at each grid index $i, j, k$, we can easily optimize performance by rapidly prototyping techniques to overlap computation and communication.
The sparsity of kernels per time-step and small number of temporary variables mean that Oceananigans' algorithmic structure is markedly different from current ocean models, which typically allocate 10 to 100 \textit{times} the minimum necessary memory \cite{balaji2017cpmip} and distribute computations across many small kernels \cite{zhang2020optimizing}.
We argue these algorithmic differences are a major factor in Oceananigans' energy-efficiency and time-to-solution on GPU systems.



\subsection{New numerical methods for finite volume fluid dynamics on the sphere}
\label{sec:num_methods}

Our results use \texttt{Oceananigans.HydrostaticFreeSurfaceModel}, which solves the hydrostatic Boussinesq equations in a finite volume framework on staggered C-grids \cite{arakawa1977computational}.
Oceananigans' hydrostatic model employs an implicit-explicit second-order Adams-Bashforth time stepping scheme.
Vertically-implicit diffusion is implemented with a backward Euler time-discretization and tridiagonal solver.

A major innovation is a new adaptive-order scheme based on weighted essentially non-oscillatory (WENO) reconstructions \cite{weno} for advecting momentum and tracers on curvilinear finite-volume grids \cite{wenopaper}.
This new scheme automatically adapts to changing spatial resolution and permits stable, high-fidelity simulations of ocean turbulence without explicit dissipation or hyper-dissipation.
This innovation reduces setup time when changing or increasing resolution while guaranteeing high-fidelity solutions that exhibit the minimum necessary dissipation of sharp, near-grid scale features.

\begin{figure}[bhtp]
  \begin{center}
    \includegraphics[width=\textwidth]{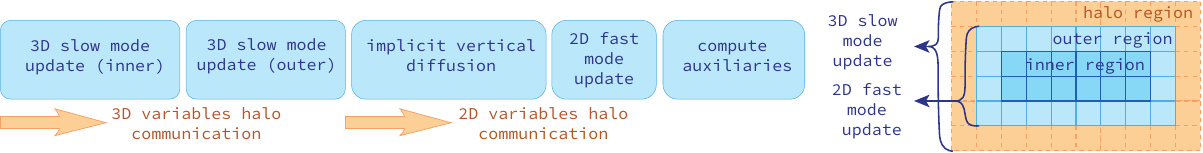}
  \end{center}
  \caption{\footnotesize Left: time-stepping sequence. Right: different domains over which 2D fast and 3D slow mode updates take place (here assuming 1 barotropic substep per baroclinic step -- halo region of size 1 -- and second-order methods -- outer region of size 1)}
  \label{fig:time-step}
\end{figure}
\normalsize

\subsection{Optimization of ocean free surface dynamics for unprecedented GPU scalability}

In hydrostatic ocean models with a free surface, the vertically-averaged, two-dimensional ``barotropic mode'' has dynamics orders of magnitude faster than the three-dimensional ``baroclinic'' component, and must be treated by a special ``barotropic solver''.
Due to communication overhead, barotropic solvers in current ocean models --- whether implicit or explicit --- are a major bottleneck that accounts for between 40\% \cite{veros} to 60\% \cite{licom3, mpas} of the cost of a typical IPCC-class ocean simulations.

Oceananigans' excellent scalability is enabled by an innovative optimization of the parallel barotropic solver. 
An increase in computation is traded in for decreased communication latency by leveraging the two-dimensionality of the barotropic problem.
Our new barotropic solver is based on explicit subcycling of the barotropic mode.
Increasing the width of the barotropic halo to equal the number of explicit subcycles (typically between 10--30) greatly decreases the frequency of communication.
As a result, communication is required once per time-step rather than every subcycle, reducing the frequency of communication by a factor of 10 to 30.
The cost of the barotropic solver is therefore less than 10\% of the total cost of a time step.
Due to the sparsity of communication enabled by our novel barotropic solver, all communication operations can be overlapped with computational workloads as sketched in figure~\ref{fig:time-step}.

\section{How performance was measured}
\label{sec:model-configuration}

\newcommand{\Otwelve}       {{\bf OceananigansR12}}
\newcommand{\Otwentyfour}   {{\bf OceananigansR24}}
\newcommand{\Ofortyeight}   {{\bf OceananigansR48}}
\newcommand{\AquaPlanet}    {{\bf OceananigansAP}}

The Oceananigans model performance is estimated for two near-global ocean simulations with different domains: a realistic (R) domain and an aqua planet (AP) domain.
Both domains span the entire longitudinal extent of the sphere and cover a latitude range of $75^\circ$S to $75^\circ$N. 

\begin{figure}[htp]
    \centering
    \makebox[\textwidth][c]{\includegraphics[width=1.0\textwidth]{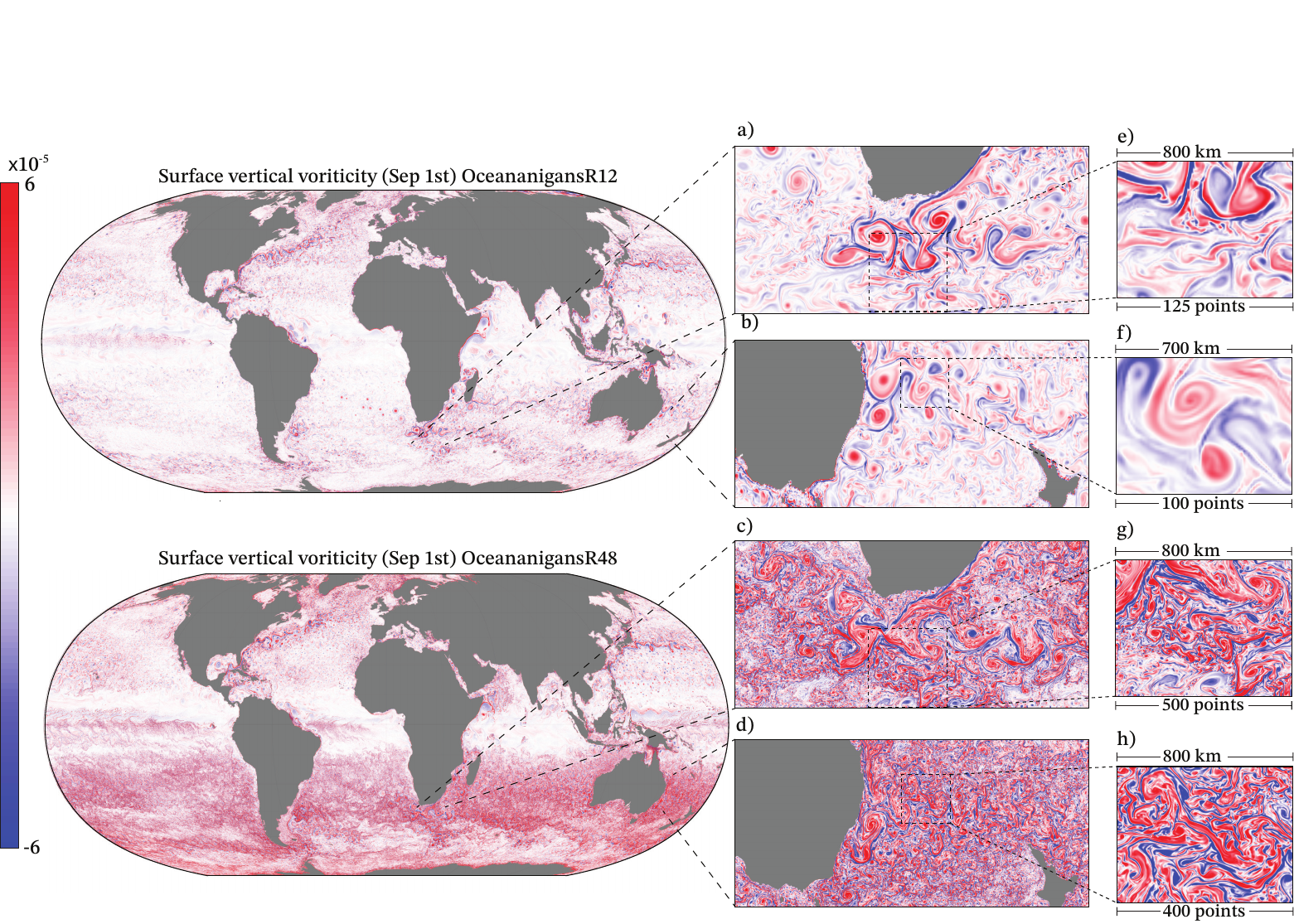}}
    \caption{ \footnotesize Vertical vorticity as simulated by {\bf Oceananigans12} (top left) and {\bf Oceananigans48} (bottom left) after a one year integration  on September 1st.
    To the right, insets zoom on particularly energetic current systems: the Aghulas and the East Australian Currents.
    While major ocean currents with widths of 10-100 km are resolved in both simulations, the sharp density fronts and associated currents that develop at the ocean surface in winter at scales between 1-10 km (the ocean weather) are only resolved by {\bf Oceananigans48}.
    On September 1 --- spring in the southern hemisphere, fall in the northern hemisphere --- such sharp frontal features populate the southern ocean but are suppressed in the north.}
    \label{fig:submesoscale-ocean}
\end{figure}

The Realistic domain has realistic bathymetry and is forced by realistic surface momentum, heat, and salinity fluxes derived from the ECCO2 state estimate\cite{menemenlis2008ecco2} at three resolutions:
\begin{itemize}
\item \Otwelve \,     with 1/12${}^{\rm{th}}$ degree horizontal resolution ($\sim$7 km)   and 48 vertical levels
\item \Otwentyfour \, with 1/24${}^{\rm{th}}$ degree horizontal resolution ($\sim$3.4 km) and 100 vertical levels
\item \Ofortyeight \, with 1/48${}^{\rm{th}}$ degree horizontal resolution ($\sim$1.7 km) and 100 vertical levels 
\end{itemize}
Figure~\ref{fig:submesoscale-ocean} shows surface vertical vorticity after one year integration of \Otwelve \, and \Ofortyeight \, over the global ocean and also for selected regions to show further detail.
Both \Ofortyeight \, and \Otwelve \, exhibit macroscale turbulent ocean features that are currently unresolved by most IPCC-class models.
The \Ofortyeight \, solution exhibits fronts, filaments, and other ``submesoscale'' vorticity features realized only a handful of times in global simulations.

The idealized \AquaPlanet{} suite of simulations \cite{doubledrake}, which has idealized bathymetry and surface forcing that does not require interpolation to different resolutions, is used for weak scaling experiments.
All \AquaPlanet{} experiments have 100 vertical levels and two latitudinal ridges that divide the world ocean into two basins.
We vary the horizontal resolution of \AquaPlanet{} from 1/6${}^{\rm{th}}$ of a degree ($\sim$14$\,$km) to 1/196${}^{\rm{th}}$ of a degree ($\sim$488$\,$m). 

None of our simulations require explicit horizontal diffusion of momentum or tracers owing to the adaptive WENO advection scheme described in section~\ref{sec:num_methods}.
All simulations use a Richardson-number-based parameterization for vertical mixing due to unresolved shear and convective turbulence at 1--100$\,$m scales.

To assess the time-to-solution for each experiment in simulated years per day (SYPD), we measure average wall clock time per time-step.
Wall clock time is sampled through NVIDIA's Nsight System and recorded by NVIDIA Tool Extension Library via the NVTX.jl Julia package.

To assess the efficiency of each solution in simulated years per mega-watt-hour (SYPMWh), we combine SYPD with an estimate of the mean power draw over the duration of an experiment.
On MIT Satori~\cite{mit-satori}, which has 256 Nvidia V100s, we have access to precise, billing-grade power metering.
For all simulations with Nvidia A100s we estimate power consumption $P$ with
\begin{equation} \label{a100-power}
P = 250 D + 300 N \, \text{Watts} \, ,
\end{equation}
where $D$ is the number of A100s and $N$ is the number of nodes.

We further note that power estimates are provided by LICOM3 and COSMO, but not for LLC4320 or Veros.
To estimate the power consumption of LLC4320, we assume that each of the 1000 dual CPU nodes draws 500W.
We estimate the power consumption of iHESP CESM \cite{zhang2020optimizing} and HadGCM3~\cite{highresmip} as a percentage of the peak power consumption of their respective clusters.
We use equation~\eqref{a100-power} to estimate Veros' power consumption on 1 node with 16 A100s.

\normalsize

\section{Performance Results}

We report both scaling results via time-to-solution in SYPD and efficiency results via energy-to-solution in SYPMWh.

\subsection{Scaling Results}

\begin{figure}[htp]
  \begin{center}
    \includegraphics[width=\textwidth]{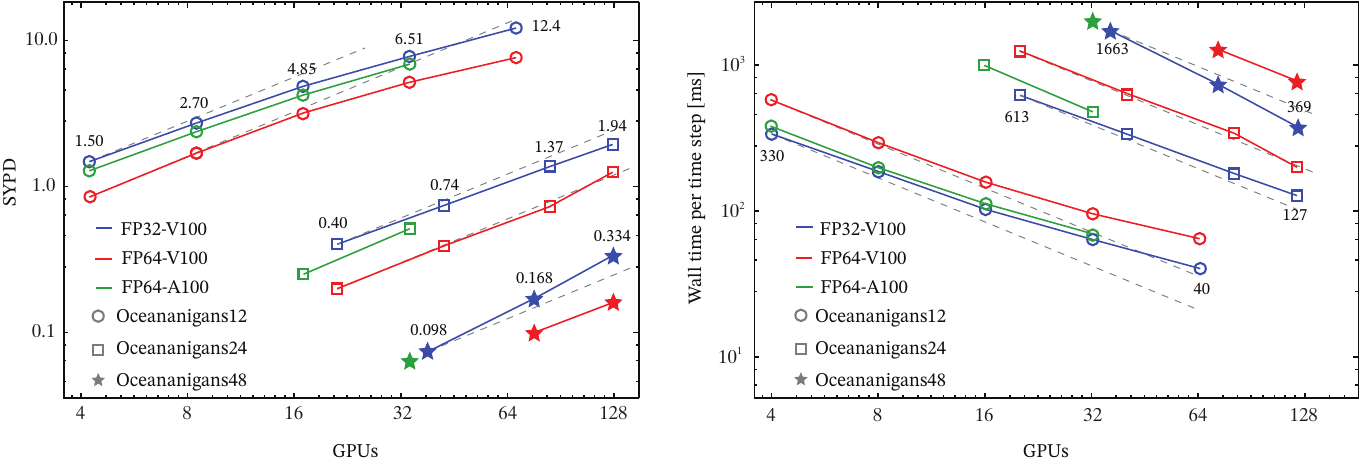}
  \end{center}
  \footnotesize
  \caption{\footnotesize Strong scaling tests for the realistic setups \Otwelve \, ($1/12^\circ$), \Otwentyfour \, ($1/24^\circ$), and \Ofortyeight \, ($1/48^\circ$).
  The left plot reports simulated years per wall clock day (SYPD) while the right plot wall clock milliseconds per time steps. All results are averaged over 1500 time steps.}
  \normalsize
  \label{fig:strong-scaling}
\end{figure}

\normalsize

\textbf{Realistic ocean simulations (Satori and Engaging clusters).}
We report strong scaling tests using the realistic global setup shown in figure \ref{fig:submesoscale-ocean} on two clusters: \textit{(i)} the MIT Satori cluster \cite{mit-satori}, a high-performance Power 9 system composed of 64 Power 9 nodes hosting four Nvidia V100 GPUs with 32GBs memory each, and \textit{(ii)} the Engaging MIT cluster, using 8 nodes that host 4 NVlinked A100s with 80GBs memory each.
The resulting wall clock time per time step, averaged over 1500 time steps, is presented in Figure \ref{fig:strong-scaling} for both single precision (FP32) and double precision (FP64) computations.
On a single node, \Otwelve \, attains 0.9 SYPD in double precision and 1.4 SYPD in single precision, with a wall clock time per time step ranging from 330 to 550 ms. 
When increasing the number of nodes up to 16 (64 GPUs), the communication overhead increases, resulting in 12.4 SYPD in single precision and 7.75 SYPD in double precision. 
We measure a strong scaling efficiency of 52\% in single precision and 55\% in double precision over 64 GPUs, because the computational workload (40 ms wall clock time per time-step) eventually becomes too short to completely mask the communication overhead.

\begin{figure}
  \begin{center}
    \includegraphics[width=\textwidth]{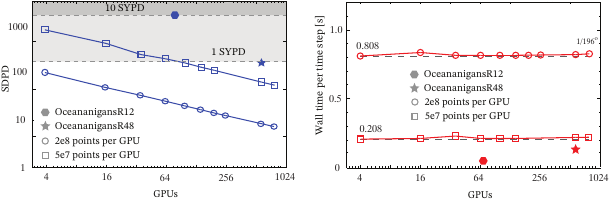}
  \end{center}
  \caption{\footnotesize Weak scaling tests performed in double precision with the {\AquaPlanet} setup.
  Each GPU has a grid equivalent to a global $1/6^\circ$ and 100 vertical layers.
  The weak scaling is performed up to a horizontal resolution of 1/168${}^{\rm{th}}$ of a degree ($\sim$488~m resolution) where we achieve 15 simulated days per wall clock day (1 year in roughly 25 days).
  The star marks the performance of \Ofortyeight \, (figure~\ref{fig:submesoscale-ocean}) on 144 Perlmutter GPU nodes.
  All results are averaged over 500 time steps.}
  \label{fig:weak-scaling}
\end{figure}

For higher-resolution ocean weather-permitting simulations, the scaling is almost ideal across the range we investigate.
For \Otwentyfour \, (FP64-V100) and \Ofortyeight \, (FP32-V100), we measure larger than ideal scaling. 
This counter-intuitive result is a product of a load balance improvement as the number of GPUs increases. 
In summary, we attain 1.94 SYPD on 120 V100 GPUs with a kilometer-scale resolution (\Otwentyfour) and 0.33 SYPD with an ocean weather-resolving simulation (\Ofortyeight). 
Finally, we have tested the \Ofortyeight \, setup on 144 Perlmutter nodes (576 A100 GPUs), reaching the 0.95 SYPD.
This is the \textit{first instance} of a kilometer-scale ocean achieving $\sim$1~SYPD.
We have also tested the \Otwelve \, setup on 17 nodes obtaining 9.9 SYPD (see fig. \ref{fig:weak-scaling}). 

\textbf{Aqua-planet simulation (Perlmutter cluster).} We report weak scaling tests on the NERSC supercomputer (Perlmutter).
Perlmutter is a HPE (Hewlett Packard Enterprise) Cray EX supercomputer that hosts four A100 GPUs with 40GB per node, linked through a NVLink3 interconnect. 
All weak scaling tests are performed using the {\AquaPlanet} setup on double precision.
We allocate two different horizontal resolutions (1/12 and 1/6 of a degree), progressively increasing them with the number of GPUs while maintaining 100 vertical levels. 
As shown in figure \ref{fig:weak-scaling}, we obtain  100\% weak scaling efficiency for the whole investigated range (1 to 196 nodes -- 4 to 768 A100s).

\subsection{Energy efficiency}

In table \ref{tab:gpu-models} we summarize the energy metrics for our computations as well as the other investigated models. Figure \ref{fig:sypmwh} is derived from the data outlined in this table.
HadGEM3 and iHRES  entries are estimated by including the whole coupled climate model (atmosphere and ocean).
Unavailable data is marked with $-$.
Our Oceananigans simulations are the highest in each of their columns.
This reflects our attention to memory and energy efficiency.

\begin{table}[htp]
\centering
\scalebox{0.75}{\begin{tabular}{lrrrrrrl}
\hline
Model & Time step & Grid size & CPU/GPU & SYPD & wtime/tstep & Power est. & In fig \ref{fig:sypmwh}\\ 
\hline
HadGEM3$_{\text{FP64}}$ (Climate) \cite{highresmip} & - & $\sim7.22\times 10^8$ & 9396 (Cray XC40) & 0.4 & - & 141KW &  $\usym{1F5F8}$\\
iHRES$_{\text{FP64}}$  (Climate) \cite{zhang2020optimizing} & - & $\sim6\times10^8$ & Sunway TaihuLight & 3.7 & - & 6500KW & $\usym{1F5F8}$\\
LLC4320$_{\text{FP64}}$  (Ocean)           & 25 s  & $8.7\times 10^{10}$   & 2000  (Intel) & 0.041 & 1.6   & 500KW &  $\usym{1F5F8}$\\
Veros$_{\text{FP64}}$   (Ocean) \cite{veros}   & 180 s & $3.5 \times 10^{8} $  & 16    (A100)  & 0.8   & 0.62  & 5.2KW & $\usym{1F5F8}$\\ 
Veros$_{\text{FP32}}$   (Ocean) \cite{veros}   & 180 s & $3.5 \times 10^{8} $  & 16    (A100)  & 1.3   & 0.38  & 5.2KW & \\ 
LICOM3$_{\text{FP64}}$  (Ocean) \cite{licom3}  & 60 s  & $1.5\times 10^{9}$    & 384   (MI50)  & 0.51  & 0.32  & 92KW & $\usym{1F5F8}$\\
LICOM3$_{\text{FP64}}$  (Ocean) \cite{licom3}  & 60 s  & $1.5\times 10^{9}$    & 26200 (MI50)  & 2.72  & 0.06  & 6300KW & \\
COSMO$_{\text{FP64}}$   (Atmos) \cite{cosmo}   & 6 s   & $3.46\times 10^{10}$  & 4888  (P100)  & 0.043 & 0.4   & 1000KW & $\usym{1F5F8}$\\
\Otwelve$_{\text{FP32}}$  \, (Ocean)              & 180 s & $3.7\times 10^8$      & 4     (V100)  & 1.5   & 0.33  & 1.2KW &$\usym{1F5F8}$\\
\Otwelve$_{\text{FP32}}$  \, (Ocean)              & 180 s & $3.7\times 10^8$      & 64    (V100)  & 12.4  & 0.04  & 18KW & \\
\Otwelve$_{\text{FP64}}$ \, (Ocean)              & 180 s & $3.7\times 10^8$      & 68    (A100)  & 9.9 & 0.05  & 22 & $\usym{1F5F8}$\\
\Ofortyeight$_{\text{FP32}}$  \, (Ocean)          & 45 s  & $1.24\times 10^{10}$  & 120   (V100)  & 0.33  & 0.37  & 36KW & \\
\Ofortyeight$_{\text{FP64}}$  \, (Ocean)          & 45 s  & $1.24\times 10^{10}$  & 576   (A100)  & 1.0   & 0.13  & 187KW & $\usym{1F5F8}$\\
\Ofortyeight$_{\text{FP64}}$  \, (Ocean)          & 45 s  & $1.24\times 10^{10}$  & 32    (A100)  & 0.063 & 1.9 & 10.4KW & \\
\AquaPlanet$_{\text{FP64}}$  \, (Ocean)          & 11 s  & $2.1\times 10^{11}$   & 768   (A100)  & 0.063 & 0.81  & 252KW & $\usym{1F5F8}$\\
\hline
\end{tabular}}
\vspace{0.3cm}
\caption{\label{tab:gpu-models} \footnotesize Performance details of state-of-the are climate, ocean, and atmosphere models. Larger grid sizes correspond to finer spatial resolution. Computations belonging to this submission are shown in bold.}
\end{table}

\section{Implications}

By developing a new model from scratch specifically for GPUs, and wielding a handful of key ocean-model-specific innovations, Oceananigans achieves 9.9~SYPD at 10 km resolution using less than 1\% of the resources of current state of the art supercomputers.
This achievement means that most climate model runs submitted to IPCC will be able use 10 km ocean models --- \textit{precipitating a step change in the accuracy of climate prediction.}

At scales between 10--100~km, macroscale ocean turbulence exerts a key control on ocean carbon and heat uptake.
However, attempts to accurately parameterize this key process in coarse resolution models have frustrated generations of oceanographers.
The inadequacies of macroscale parameterizations are associated with major biases and uncertainty in climate predictions~\cite{Mundayetal14, Saenkoetal18}.
At resolutions of 10~km, the need for macroscale turbulence parameterization is eliminated, and ocean simulations capture key ocean features such as sharp sea surface temperature gradients supporting the formation of marine stratus clouds above narrow eastern boundary currents like the California and Benguela Current \cite{ma2019warm}, and changes in the meridional overturning circulation due to the effect of Antarctic meltwater on deep convection in austral winter~\cite{li2023abyssal}.

Additionally, by achieving 0.95 SYPD at 1.7 km resolution, we pave the way for decadal ocean simulations of the ocean ``submesoscale'' --- the ocean analogue to atmospheric weather --- which exhibits hourly fluctuations, high spatial and seasonal variability, and which exerts a strong control on ocean air-sea fluxes, biological productivity and fish stocks~\cite{TaylorThompson23}.
The granularity and accuracy provided by 1.7 km resolution is further required to plan local mitigation strategies and predict local extreme events.

Third, the unparalleled speed of execution and memory efficiency of Oceananigans allows global computations at never-before-seen sub-kilometer resolutions.
The capacity for ultra-high-resolution simulations aligns with current advancements in resolution of ocean sampling platforms from satellites~\cite{swot,donlon2012sentinel} to fleets of floats and drones.
While this wealth of data is likely to provide new insights and scientific knowledge about the nature of small scale processes, global high-resolution ocean simulations will be needed to explore their impact on global climate scales.

Finally, our results pave the way for marked increase in energy efficiency of climate simulations. The very reason to develop climate models, as stated by the Coupled Model Intercomparison Project (CMIP), for example, is to provide the necessary information to effectively reduce emissions and mitigate the effects of global warming --- while, counterproductively, the carbon footprint of climate simulations that contribute to CMIP increases rapidly.
Oceananigans' achievements represent a milestone towards decreased energy consumption by climate modeling efforts.

\section{Acknowledgments}
This research used resources of the National Energy Research Scientific Computing Center (NERSC), a U.S. Department of Energy Office of Science User Facility located at Lawrence Berkeley National Laboratory, operated under Contract No. DE-AC02-05CH11231 using NERSC award DDR-ERCAP0025591.
This work is partly supported by the generosity of Eric and Wendy Schmidt by recommendation of the Schmidt Futures program and by NSF grant AGS-1835576.
N.C.C.~is supported by the Australian Research Council DECRA Fellowship DE210100749.

\bibliographystyle{plain} 

\end{document}